\documentclass[12pt]{article}
\usepackage{hyperref}
\title{Open Science, Public Engagement and the University \footnote{This is a white paper commissioned by the NSF and NIH funded conference: \url{http://www.ncsa.illinois.edu/Conferences/ImagineU/}. Correspondence can be sent to pratim.sengupta@ucalgary.ca or mcshanah@ucalgary.ca.}}
\author{
        Pratim Sengupta \footnote{Pratim Sengupta is the Research Chair of STEM Education at the University of Calgary, Canada, and a recipient of the NSF CAREER Award (2012).} \\ Marie-Claire Shanahan \footnote{Marie-Claire Shanahan is Associate Professor of Learning Sciences, and Research Chair of Science Education and Public Engagement (2013 - 2016) at the University of Calgary, Canada.} \\
        University of Calgary, Canada\\
}
\date{}

\begin{document}
\maketitle

\begin{abstract}
Contemporary debates on ``open science" mostly focus on the public accessibility of the \textit{products} of scientific and academic work. In contrast, this paper presents arguments for ``opening" \textit{the ongoing work of science}. That is, this paper is an invitation to rethink the university with an eye toward engaging the public in the dynamic, conceptual and representational work involved in creating scientific knowledge. To this end, we posit that \textit{public computing} spaces, a genre of open-ended, public learning environment where visitors interact with open source computing platforms to directly access, modify and create complex and authentic scientific work, can serve as a possible model of ``open science" in the university.
\end{abstract}

\section{Introduction}
The goal of this paper is to provoke a conversation about the nature of knowledge and the place of the university in the public realm. Contemporary debates on ``open science" mostly focus on the public accessibility of the \textit{products} of scientific and academic work. In contrast, this paper presents arguments for ``opening" \textit{the ongoing work of science}. That is, this paper is an invitation to rethink the university with an eye toward engaging the public in the dynamic, conceptual and representational work involved in creating scientific knowledge. 
\par The arguments presented in this paper are grounded in the sociological, historical and philosophical studies of science and scientists ``in action", and  in educational research on making scientific computing accessible to the public. The review presented in Section~\ref{os}  reveals key characteristics of the nature of scientific work and knowledge. This helps us identify elements of scientific work that can be pivotal points of public engagement. Section~\ref{fis} presents an argument for viewing particular forms of computing as more amenable to public engagement with computational science. In Section~\ref{results}, we present an illustrative example of public education and research that can serve as a model for ``open science" in the university. We present \textit{public computing spaces} as a form of open-ended, public learning environment, in which members of the public can interact with open source computing platforms to directly access, modify and create complex and authentic scientific work.  

\section{Definitional Issues}\label{previous work}

\subsection{The Tension Between Science and the Public}\label{os}
In our view, the real power of \textit{open science} is the potential creation of diverse opportunities for productive and meaningful engagement of the public \textit{in} and \textit{with} scientific work. But what does it mean to ``do" science? Scholars who study scientists in action generally agree that a productive way to characterize the work of science is as the iterative development and refinement of models.\footnote{Pickering, A. (1992). \textit{Science as practice and culture.} University of Chicago Press.}\footnote{Lehrer, R. (2009). Designing to develop disciplinary dispositions: modeling natural systems. \textit{American Psychologist}, 64(8), 759.}  Perhaps the most well-known proponent of this idea is the philosopher Thomas Kuhn, who argued that collectively, scientific work can be understood in terms of \textit{paradigms} of theories and models that are put forward by communities of scientists in order to explain the world.
\par For Kuhn\footnote{Kuhn, T. S. (1974). Second thoughts on paradigms. \textit{The Structure of Scientific Theories}, 2, 459-482.}, a paradigm is a ``disciplinary matrix" of symbolic generalizations, models, and exemplars that are shared by the members of a scientific community. In this paper, we will simply refer to these elements using the term ``models". However, Kuhn also pointed out that thinking of science as ``paradigmatic" also necessitates us recognizing that the scientific communities have independent existence. He wrote: 
\begin{quote}
``A paradigm is what the members of a scientific community, and they alone, share. Conversely, it is their possession of a common paradigm that constitutes a scientific community of a group of otherwise disparate men. [\ldots.] If the term “paradigm” is to be successfully explicated, scientific communities must first be recognized as having an independent existence." 	
\end{quote}
\par For Kuhn, the \textit{independence} of scientific communities is partly from one another. But perhaps most significantly for our purposes, the emphasis on independence delineates the location of scientists from the public realm. The scientist and the public remain separated by the disciplinary matrix, which can only be developed and deciphered by the scientist.
\par If science is truly open, however, the community that defines and shares a Kuhnian paradigm must also include non-scientists. In attempting to understand and describe his own practice as a chemist, Michael Polanyi reflected deeply on the mutual authority that characterizes scientific knowledge creation \footnote{Polanyi, M. (1962). Tacit knowing: Its bearing on some problems of philosophy. \textit{Reviews of Modern Physics}, 34(4), 601.}. Within scientific communities, science cannot be a strictly understood as a hierarchical practice where small individual contribution are sent up the chain of command to be combined. Nor can it be seen as a collective but solitary effort, where individuals work separately and pool their work in some sort of summative way. The everyday metaphor he found most satisfying was that of team working together on a jigsaw puzzle.
\begin{quote}
  ``The only way the assistants can effectively co-operate, and surpass by far what any single one of them could do, is to let them work on putting the puzzle together in sight of the others so that every time a piece of it is fitted in by one helper, all the others will immediately watch out for the next step that becomes possible in consequence." (Polanyi, 1962, p. 55).
\end{quote}
 
\par So, how can the disciplinary matrix of science be ``opened up" in meaningful manner for public engagement? That is, how can we support scientific engagement \textit{in} public and \textit{for} public purposes? In using the word public we are careful to engage the term primarily as a description of experience in a space and its materiality rather than only as a noun for those involved. It is in this latter sense that the term public typically appears in arguments for “public engagement”\footnote{See for example: Leshner, A. I. (2003). Public engagement with science. \textit{Science}, 299(5609), 977-977.}. As we have argued elsewhere\footnotemark[14], even with the recognition that there can be no single ``public" in relation to science and technology, the very noun itself is colonized with images of passive or resistant recipients of finalized knowledge. It is this conception of public that we confront here, by challenging the image of academic knowledge as a \textit{private language} - a secret code that can only be known by an individual or a chosen few. In what follows, we argue that \textit{computational science} can serve as a suitable \textit{mode} of public engagement with the private language of science.             

\subsection{Computational Science as a Potential Mode of Public Engagement}\label{fis}
 
\par Computation is inherently reflexive with science. While the notion of computational thinking\footnote{Wing, J. M. (2006). Computational thinking. \textit{Communications of the ACM}, 49(3), 33-35.} involves being able to \textit{think} in terms of disciplinary lenses central to computer science (for example, data structures, algorithms, etc.), it also includes \textit{practices} such as decomposition, simulation, verification, and prediction. These practices are both epistemic and representational in nature.\footnote{For a detailed discussion and review, see: Sengupta, P., Kinnebrew, J. S., Basu, S., Biswas, G., \& Clark, D. (2013). Integrating computational thinking with K-12 science education using agent-based computation: A theoretical framework. \textit{Education and Information Technologies}, 18(2), 351-380.} That is, they are involve the simultaneous development of ideas and design of external representations. Working on one aspect cannot proceed without working on the other. These practices, in turn, are also central to modeling, reasoning and problem solving in a large number of scientific, engineering and mathematical disciplines.\footnotemark[7]
\par Educational researchers have identified some interesting synergies between computational modeling and programming on one hand, and the development of scientific understanding in K-12 students on the other. For example learners' conceptual difficulties in math, science and programming exhibit similar patterns. Conceptual difficulties in each of these domains have both domain-specific roots (e.g., challenging concepts) and domain general roots (e.g., difficulties pertaining to conducting inquiry, problem solving, and epistemological knowledge). Complementarily, learning programming in concert with concepts from another domain can be easier than learning each separately\footnote{Harel, I., \& Papert, S. (1990). Software design as a learning environment. \textit{Interactive Learning Environments}, 1(1), 1-32.}. Several researchers have shown that programming and computational modeling can serve as effective vehicles for learning challenging science and math concepts and, further, that learning through computation can contribute to re-aligning the practices of science education with those of scientific research\footnotemark[7] \footnotemark[10].
\par Computing education research has also yielded advances in terms of identifying \textit{agent-based computing} as a genre of computing that are be effectively utilized for science education, even for young learners (e.g., elementary students)\footnote{Dickes, A. C., Sengupta, P., Farris, A. V., \& Basu, S. (2016). Development of Mechanistic Reasoning and Multilevel Explanations of Ecology in Third Grade Using Agent‐Based Models. \textit{Science Education}, 100(4), 734-776.} \footnote{Sengupta, P., Dickes, A., Farris, A. V., Karan, A., Martin, D., \& Wright, M. (2015). Programming in K-12 science classrooms. \textit{Communications of the ACM}, 58(11), 33-35.}. In the agent-based paradigm, the user programs the behaviors of one or more virtual agents by using simple computational rules, which are then executed or simulated in steps over time to generate an evolving set of behaviors. Research shows that when students learn using agent-based models and simulations, they first use their intuitive knowledge at the agent level to manipulate and reason about the behaviors of individual agents. As they visualize and analyze the aggregate-level behaviors that are dynamically displayed in the agent-based simulation environment, students can gradually develop multi-level explanations by connecting their relevant agent-level intuitions with the emergent aggregate-level phenomena\footnotemark[9] \footnote{Wilensky, U., \& Reisman, K. (2006) Thinking Like a Wolf, a Sheep, or a Firefly: Learning Biology Through Constructing and Testing Computational Theories—An Embodied Modeling Approach, \textit{Cognition and Instruction}, 24:2, 171-209}. Such pedagogical approaches enable children to develop deep explanations of complex scientific phenomena by building upon, rather than discarding their repertoire of intuitive knowledge. Furthermore, technological innovations include the development of high-ceiling and low-threshold, open-source programming languages such as NetLogo\footnote{Tisue, S., \& Wilensky, U. (2004, May). Netlogo: A simple environment for modeling complexity. In: \textit{Proceedings of the International Conference on Complex Systems}, Vol. 21, pp. 16-21.} and Processing\footnote{Reas, C., \& Fry, B. (2007). \textit{Processing: a programming handbook for visual designers and artists} (No. 6812). MIT Press.}, that are both usable by both beginners and experts.
\par We therefore believe that computational science can become a mode of public engagement with the disciplinary matrix of science. In what follows, we present a brief description of such an effort that is currently underway at the University of Calgary's Werklund School of Education.

\section{An Illustrative Example: Public Computing}\label{results}
In a recent paper, we paradigmatically argued for a frame shift in the technological infrastructure as it pertains to computationally intensive science and public education\footnote{Sengupta, P., \& Shanahan, M.C. (In Press). Boundary Play and Pivots in Public Computation: New Directions in STEM Education. \textit{International Journal of Engineering Education.}}. We introduced \textit{public computing spaces} as a form of open-ended, public learning environment, in which visitors interact with open source computing platforms to directly access, modify and create complex and authentic scientific work. 
\par In formal education, the meaning of individual scientific disciplines is often formed in reference or opposition to codified forms of disciplinary knowledge and culture such as curricula, textbooks and accepted teaching and learning models. These can form a rigid paradigm for learners, limiting the interactions they may have with each other, with researchers and with scientific knowledge itself.\footnote{Cornelius, L. L., \& Herrenkohl, L. R. (2004) Power in the classroom: How the classroom environment shapes students' relationships with each other and with concepts. \textit{Cognition and Instruction}, 22(4), 467-498.} Outside of formalized learning environments, public access and interactions with scientific research and researchers are sometimes similarly constrained by paradigms and histories of public understanding of science that construct the public as largely passive recipients of scientific knowledge, much like students in classrooms \footnote{Wynne, B. (2006). Public engagement as a means of restoring public trust in science–hitting the notes, but missing the music?. \textit{Public Health Genomics}, 9(3), 211-220.}. 
\par By giving learners unrestricted access to the matrix of scientific computation - the code itself - public computing spaces can offer opportunities, technological means and human capital to play with disciplinary meanings and expertise in authentic, yet novel and unexpected ways. Open source computing can further facilitate this process by opening up the “code”, which often reifies epistemic and representational work of experts, for the public. The opening up of epistemic and representational possibilities, we argue, are both due the structural affordances of the computing media (e.g., open source and sensors-based interactivity), as well as the opportunities of collaboration with friends, strangers and experts that often get taken up through joint action, as users configure and reconfigure novel scientific representations and their explanations. 
\par To this end, we developed DigiPlay, a public learning environment that uses open source computing for STEM experiences \textit{in} and \textit{for public}. DigiPlay is a learning environment located in an indoor, public walkway at the University of Calgary. It consists of three 80" touch screens, each powered by a desktop. The screens currently display open source simulations of complex systems. Visitors can use the touch-sensitive screens to interact with simulated visualizations of complex systems in which the larger scale patterns (flocks) emerge as each virtual bird performs simple interactions with neighbouring birds. The simulations are programmed using the Processing programming language and visualization platform. Processing is open source, used by professional computer scientists and digital artists alike, and there is a strong online user community of experts and learners, making sample code and simulations accessible to the public.
\par The current setup in DigiPlay comprises of simulations of complexity, designed in the Processing language. These simulations are both open source and glass box. The open source nature of the code makes it possible for visitors to interact with and modify the code that may have been originally created by an expert, and it also allows us as developers of DigiPlay to extend and modify functionalities of the Processing programming language itself, as needed. The glass box nature of Processing enables that visitors to access the underlying code, while the simulations are running in the form of dynamic visualizations in fullscreen mode, by simply hitting the “Escape” button once on the on-screen keyboard. DigiPlay visitors can directly interact with and modify the emergent patterns in the visualizations by adding new boids to the flock by touching the screens, and at the same time they can  also make deeper changes to the way the individual boids interact by accessing the underlying code. 
\par The algorithms we used are adapted from Reas \& Fry's\footnotemark[13] implementation of Craig Reynold’s classic algorithm for simulating flocking of birds (Reynold termed each virtual bird a ``Boid"). Each Boid in the simulation acts as a computational “agent”, and the DigiPlay simulations can therefore can be understood as multi-agent simulations. The term “agent” here indicates individual computational objects or actors. It is the behaviors and interactions between these agents as well as elements of the environments in which they are situated, that give rise to emergent, system-level behavior (e.g., the formation and movement of a traffic jam or the spread of disease). Each agent in a multi-agent simulation makes it own decision. Therefore the emergent patterns represented in the simulations do not result from averaging over a population but from the aggregation of the outcomes of individual-level decisions of multiple agents. This concept forms the central scientific meaning of the simulations.
\par The rules obeyed by each Boid in the simulations are as follows: alignment, separation, and cohesion. Alignment means that a Boid tends to turn so that it is moving in the same direction that nearby birds are moving. Separation means that a Boid will turn to avoid another bird which gets too close. Cohesion means that a Boid will move towards other nearby birds (unless another bird is too close). The relevant portion of the code that controls the relative weights of these “rules”, is used as the pivotal code fragment by facilitators to explain to the visitors both how the simulations work in terms of the agent-level rules, as well as to provide them with opportunities to directly alter the key interactions between the Boids, and thus generate new patterns of emergent behaviors.   
\par Finally, the public nature of the space ensures that anyone can walk in and interact with DigiPlay. The users can access just-in-time information as they are interacting with the touch screens, which provide them instructions for modifying the live simulations. The “rules” obeyed by the Boids are also explained in the form of posters on glass walls surrounding the monitors.  In addition, there is often an on-site facilitator present to provide the public direct and live access to expertise. The on-site facilitator is one of the members of the team that developed the exhibit (including the first author), and they have a deep understanding of the underlying code. The facilitator’s primary role is encouraging the visitors to “hack” the simulations, by showing them how to access the underlying code, and pointing them to relevant areas in the code that can be easily altered to potentially powerful effect. The interested reader may consider for further reading our research papers\footnotemark[14] on empirical observations of creative computing by visitors across a wide age range (elementary grade children to graduate students). These papers will provide a clearer sense of the kinds of deep engagement\footnote{A nice discussion on forms of engagement can be found here: Willms, J. D., Friesen, S., \& Milton, P. (2009). What Did You Do in School Today? Transforming Classrooms through Social, Academic, and Intellectual Engagement. \textit{First National Report of the Canadian Educational Association}.} that can be fostered in such spaces.

\section{Discussion: ``Open" Science, Uncertainty and Public Experience}\label{conclusions}
\par Paradigms of public understanding of science and even public engagement with science, both assume that there can be little overlap in the epistemological roles of scientists and non-scientists. The latter group is usually meant as ``the public". Even when they contribute early in the scientific process, through public consultations for example, or become data-gathering participants in citizen science project, the public-ness of the scientific work is primarily limited to inputs and outputs of scientific knowledge. Non-scientists can contribute to driving socially relevant research directions or they can gain access to open access publication of scientific results. But the internal processes of a scientific community - Kuhn's disciplinary matrices - rarely cross the line of publicness. These input-output relationships are also often constructed around assumptions of certainty: that public concerns can be concretely answered by scientific research or that knowledge communicated to the public is certain and final.
\par The world of science, as experienced by scientists on the other hand, is much more complex. Rarely is it the case that the transformation of an initial idea to a successful scientific experiment (or model) is a simple and linear process. The philosopher Andrew Pickering\footnotemark[1] pointed out that scientists are always enmeshed in a ``mangle of practice". That is, scientists struggle continuously in order to get instruments and the natural world to perform in the ways that their investigations require. The creation of scientific knowledge can therefore be understood as a dynamical process of interactive stabilization of material and human agency. Uncertainty is therefore an unavoidable aspect in the \textit{experience} of scientific work, even when the result of this work may have the appearance of certainty. It is therefore no surprise that a big focus of the scientific work is the documentation of this \textit{mangle}, in the form of meticulous note-taking, working papers, etc., which is essential for the verifiability and reproducibility of scientific work. Studies of discourse among collaborating scientists reveal the meta-stable nature of scientific knowledge and the role that interpretive work and negotiation play in dealing with uncertainty \textit{during} a research project\footnote{Ochs, E., Gonzales, P., \& Jacoby, S. (1996). ``When I come down I'm in the domain state": Grammar and graphic representation in the interpretive activity of physicists. \textit{Studies in Interactional Sociolinguistics}, 13, 328-369.}.
\par The image of certainty that is often popularly associated with scientific knowledge, therefore, is an inaccurate representation of the work of scientists. Our concern is that the emphasis on open-access of published scientific work\footnote{For an excellent discussion on open access publishing, please see: Willinsky, J. (2006). \textit{The access principle: The case for open access to research and scholarship.} Cambridge, Mass.: MIT Press.}, while necessary, may not be sufficient to address this epistemological divide, and instead may further reify the popular yet grossly inaccurate notion of certainty of scientific knowledge in the public realm. To this end, we believe that our proposal of public computational spaces can serve as \textit{a} (not \textit{the}) model solution. In \textit{Democracy \& Education}, John Dewey poignantly noted that one cannot engage in public life all by oneself. This means that the public engagement in science must also be designed or supported, both from the perspectives of the scientist as well as the public (or non-scientists). This is the work that the university must undertake for deep and widespread public engagement with science. Making public the products of science may not provide sufficient support to the public to engage with science. Our ongoing work on public computation, on the other hand, suggests that our visitors - many of whom are strangers to the work of science - can engage creatively (i.e., in a personally meaningful manner) as well as authentically (i.e., in a scientifically meaningful manner) with scientific work\footnotemark[14].  
\par Public computation spaces can be easily imagined and established \textit{in} the university, and such spaces can engage non-scientists directly with the work of scientists through open-source scientific simulations. Similar spaces can also be easily imagined, where academics from other disciplines can work alongside non-academics and co-create artifacts of value for the academic community. Such work does not intend to be an experience ``for the public" in the sense of a lesson taught explicitly to outsiders. In contrast, these examples are models of scientific experiences “in the interest of publicness”, where scientific (and more broadly, academic) work - not merely its products - can become public experience. 
\paragraph{Acknowledgements\\}
The authors would like to acknowledge Dennis Sumara and Sharon Friesen for their leadership and support that made this work possible. Financial support from the National Science Foundation (US), National Institute of Health (US) and Imperial Oil Foundation (Canada) are gratefully acknowledged. All opinions expressed here are the authors', and are not shared by the funding agencies.


\end{document}